\documentstyle[preprint,aps]{revtex}
\tightenlines
\def\Journal#1#2#3#4{{#1} {\bf #2}, #3 (#4)}

\def\AP{{\em Ann. Phys.}}
\def\CMP{{\em Commun. Math. Phys.}}
\def\IJMPB{{\em Int. J. Mod. Phys.} B}
\def\JETP{{\em Sov. Phys. J. JETP}}

\def\NPB{{\em Nucl. Phys.} B}

\def\PL{{\em Phys. Lett.}}

\def\PLB{{\em Phys. Lett.} B}
\def\PM{{\em Philos. Mag.}}
\def\PRL{\em Phys. Rev. Lett.}
\def\PRSLA{{\em Proc. Roy. Soc. London} A}
\def\PR{{\em Phys. Rev.}}

\def\PRB{{\em Phys. Rev.} B}
\def\PRD{{\em Phys. Rev.} D}

\def\RMP{{\em Rev. Mod. Phys.}}

\def\ZP{{\em Z. Phys.}}
\def\ZPB{{\em Z. Phys.} B}

\newcommand{\be}{\begin{equation}}
\newcommand{\ee}{\end{equation}}
\newcommand{\bea}{\begin{eqnarray}}
\newcommand{\eea}{\end{eqnarray}}

\newcommand{\hf} {{1\over2}}
\newcommand{\nonu}{\nonumber\\}
\def\la{\langle}
\def\ra{\rangle}
\def\ord{{\cal O}}
\def\eq#1{(\ref{#1})}

\begin{document}
\title{An effective theory for conductance by symmetry breaking}

\author{Sebastiao Correia$^a$\thanks{correia@lpt1.u-strasbg.fr},
Janos Polonyi$^{ab}$\thanks{polonyi@fresnel.u-strasbg.fr},
Jean Richert$^a$\thanks{richert@lpt1.u-strasbg.fr}}
\address{$^a$Laboratoire de Physique Th\'eorique\thanks{Unit\'e Mixte 
de Recherche CNRS-Universit\'e, UMR 7085},Université Louis Pasteur \\
3 rue de l'Universit\'e 67084 Strasbourg Cedex, France}
\address{$^b$Department of Atomic Physics, L. E\"otv\"os University\\
P\'azm\'any P. S\'et\'any 1/A 1117 Budapest, Hungary}

\date{\today}
\maketitle
\begin{abstract}
An effective theory is suggested for the particle-anti particle and
the particle-particle modes of strongly disordered electron systems.
The effective theory is studied in the framework of the saddle point 
expansion and found to support a vacuum which is not invariant under 
translations in imaginary time. The Goldstone bosons of this symmetry 
breaking generate a pole in the density-density correlation function. The 
condensate of the auxiliary field corresponding to the particle-particle
channel produces conductivity without relying on the long range 
fluctuations. The results are obtained for $d>1$.
\end{abstract}

\section{Introduction}
The understanding of the conductor-insulator transition of strongly
disordered systems represents 
a challenging problem due to the large number of effects which may
come into the scene. We approach this problem in this paper by means of
an effective theory description. Such a method is slightly better known
and more widely used in the High Energy Physics community 
but we hope that such a different point of view may lead to a 
more complete picture.

The description of the influence of local impurities on the collective
phenomena started with the one particle approximation. The presence
of the periodic crystal and the electron-ion interaction yields a model for 
the Bloch-Wilson band insulators \cite{blwi} and for the Peierls transition
\cite{peie}, the difference being the dynamical origin of the gap opening
at the Fermi surface. As the disorder increases such a description
becomes inapplicable and one arrives at the disorder induced
transition to Anderson localization \cite{andl}, understood 
qualitatively by simple scaling arguments \cite{scal}.
In addition to the disorder, the electron-electron interaction, as well, 
was thought to be responsible for the Mott transitions \cite{mott} where
the gap originates either in the long range order of the moments or in 
the quantum phase transition induced by charge correlations. 

After these first intuitive steps more systematic approaches
were developed along two different lines, based on the partial
resummation of the perturbation expansion and the 
non-perturbative use of effective models.

The Schwinger-Dyson resummation of the one-loop self energy for
the electron propagator or the Bethe-Salpeter resummation of the
one-loop ladder exchange reproduced the finiteness of the
lifetime of a charge on a random, static impurity background
and the classical conductivity expressions \cite{kolu}-\cite{agd}. 
The crossed exchange contributions to the conductivity \cite{lane}-\cite{glk}
and the cancellation of the diffusion pole \cite{vowo} underlined
the similarity between the conductance in strongly disordered
systems, and the long range modes in the particle-particle scattering
process. Numerous works were devoted to the introduction of the 
electron-electron interaction \cite{inte} to make the 
partially resummed perturbative description more realistic.

In the other approach different effective theories are used 
for the strongly disordered systems which are based
on the self-consistent mode-coupling and operate with
effective fields responsible for the electron propagation
\cite{wegn}-\cite{prsc}. The dynamical origin of these matrix fields 
was found in the imaginary time formalism \cite{elkh}-\cite{beki}
where the matrix index is provided by the Matsubara frequency
in the energy space. 

There are several examples where the sudden appearance of
conductivity, as the disorder is decreased, is related to the 
dynamical breakdown of the time reversal invariance. This is obvious for
weak localization where the loss of constructive
interference for the time reversed trajectories removes the
driving force to localization. The descriptions based on
effective models offer another mechanism by providing a
continuous symmetry whenever the time reversal invariance is
intact. In fact, the pole in the current correlation function
is generated in a manner reminiscent of the Goldstone theorem \cite{mcst}.
The continuous symmetry in question is the mixing of the fields
responsible for the retarded and the advanced propagators.
When the matrix valued fields in the Euclidean energy space are 
used \cite{elkh}-\cite{beki} then the retarded and the advanced
propagators are related to the positive and the negative 
Matsubara frequency components of some fields. The continuous
symmetry which is broken dynamically in the conducting phase
is the mixing of the positive and the negative
Matsubara frequency modes. It is a special rotation in the space
of the Matsubara modes which flips the sign of the frequency
and exchanges the retarded and the advanced propagators:
the time reversal transformation. 

There are two features of these effective theories which one
finds unsatisfactory:
\begin{itemize}
\item The mixing of the positive and the negative Matsubara frequency modes is
an {\bf approximate symmetry} only because the time derivative piece
of the action is clearly non-symmetrical. Even though this
term should be irrelevant at the phase transition
it would be preferable to have an exact formal symmetry whose breakdown is the 
dynamical origin of the phase transition. The origin of the exact symmetry
is hidden by the use of the Matsubara frequency space. The circumstance that the 
dimension of the Matsubara frequency space is proportional to the UV cutoff
in time suggests that the symmetry in question is actually
a time dependent gauge symmetry. In fact, the time dependent and
space independent gauge transformations mix the Matsubara modes and the time
derivative term in the action is the only one which
breaks the symmetry with respect to them. 
The spontaneous breakdown of the symmetry with respect 
to a global subgroup of a gauge group has already been
successfully treated in space-time, avoiding the use
of the momentum or the frequency space \cite{higgs}. 
Thus it is more natural to recast the effective theories
\cite{elkh}-\cite{beki} by using the time rather than the frequency variable.

\item Another difficulty related to the use of the frequency space is the
loss of the control over {\bf non-locality} of the effective theory. The 
models \cite{elkh}-\cite{beki} are based on bilocal fields and even if the
saddle point is localised the fluctuations around it correspond
to a genuinely non-local model. In lacking of the general 
theory of the non-local field theories the reliability of the
loop expansion around such a non-local background field is not known
and the apparent simplicity of the one-loop solution is misleading.
\end{itemize}

The cure of both problems is the use of time rather than frequency.
We present in this paper a simple effective theory for the phase transition 
corresponding to the appearance of conductance by the dynamical 
breakdown\footnote{The dynamical breakdown of a symmetry differs from the 
spontaneous one in what it occurs at finite instead of zero energy, and is 
identified by an order parameter with finite length or time scales as opposed to the
homogeneous order parameter of the spontaneous symmetry breaking.} of gauge 
and time inversion symmetry in space-time. 

Our basic assumption is that 
the direct effects of quenched disorder on the electrons can be
incorporated in an effective theory with quasi-local interactions in time. 
Thus the modification of the program of refs. \cite{elkh}-\cite{beki} 
consists of the use of local composite fields in time instead of the matrix
valued fields and the generation of the non-locality by quasi-local, higher order
time derivative terms in the action. The expected non-locality together with
the breakdown of the
gauge and the time reversal invariance will be generated in the framework of 
the saddle point expansion by a dynamical symmetry breaking. The non-locality
in time will originate from the presence of a ``condensate'', a non-trivial 
saddle point in time. It will be shown that a finite number of higher order derivatives
in time is sufficient to generate the desired symmetry breaking pattern. This 
description, kept in space-time, is simpler than the one given in frequency space.
Another bonus of keeping the time in the description
is the fact that the time dependent saddle point breaks an
exact continuous symmetry, the time translation symmetry
which here provides a mechanism to generate conductivity.

The important pieces of the effective theory to generate
the conducting phase, the higher order derivative terms
in time are irrelevant in the disordered,localized phase thus
their coupling constants are not yet constrained by experimental 
data collected in the localised phase. 
We do not derive the couplings either, only show that, as they depend on the 
environmental parameters, such as the Fermi energy of the 
electrons, they can drive the phase transition to delocalisation
without playing an important role in the localized phase.
The justification of our choice of the effective coupling 
constants, which we leave for a later work, can be addressed 
by means of the current methods in Quantum Field Theory \cite{erg}. 
All what is needed is a computation of the connected or the
one particle irreducible (1PI) graphs with high accuracy in 
their dependence on the external momenta in the infrared regime.

We restrict ourselves to the discussion of the effects
of quenched impurities, ignoring the Coulomb
interaction. A more reliable description naturally requires
more sophisticated methods, such as the renormalization group
treatment \cite{fink}, \cite{erg}. The failure or the success
of these methods devised for local theories will be the
measure of the importance of keeping the effective theory
local.

The organization of the paper is the following. Our effective
theory is introduced in Section II for the diffuson and the 
cooperon auxiliary fields. The symmetry breaking mechanism with
inhomogeneous vacuum is briefly introduced in Section III. The
effective model with diffusons and cooperons is discussed
in Sections IV and V, respectively. Section VI is a summary
of the work and the results.

\section{Effective Theories}
One of the most important effects to reproduce in an effective
theory is the appearance of non-local structures, such as
composite particles or condensates. If the interaction 
generated by the elimination of certain modes is of short
range and attractive then new extended bound states
may appear. If this interaction is of long range
then condensation and spontaneous symmetry breaking may take
place. We believe that the quenched disorder which is usually taken
into account by some auxiliary field does not generate
long-range interaction in space. Instead, the
dynamical symmetry breaking happening in the time direction
suggests that the static nature of the impurities leads to 
long-range interactions in time which must be 
preserved in the effective description. 
Our goal is to find an effective action which incorporates
in an economical manner this feature which we believe to be the 
key to the interactions generated by the quenched impurities.

Let us start with a remark about the distinction of 
the eventual non-locality of the interaction vertices in the action from 
those of the interactions generated by it. The locality of an
action can be classified by considering
the distance $r$ in wich the field variables are coupled
in units of the UV cutoff. The potential term without gradient
is ultra-local, $r=0$. The models with derivatives up to a finite
order, $0<r<\infty$ are quasi-local. Finally
the non-local theories are with gradients of infinite order,
$r=\infty$. The relation between this parameter $r$ of certain terms
in the action and their contribution to the correlation length $\xi$
(in units of the cutoff) reflects the relevance of the terms in question.
In fact, an irrelevance of a higher order derivative term with 
dimensionless coupling constant $g$ implies $d\xi/dg\to0$
as the cutoff tends to infinite. It is worthwhile noting that an
effective interaction which arises from the 
elimination of some local degrees of freedom is always formally
non-local. This happens because those contributions to the 1PI (1
particle irreducible)
functions where the internal lines correspond
to the eliminated modes contribute to the effective coupling 
constants induced by the blocking and they 
display non-polynomial momentum dependence. 
The corresponding blocked action, the generating functional,
is non-polynomial in the gradient. The important question is whether
the infinite series in the gradient can be truncated at a finite
order without modifying the universality class of the model.
If this can be done the effective theory can safely be classified as
quasi-local. 

The issue of non-locality of the interactions arises
at the IR end point, when all modes are 
eliminated and the true 1PI functions are considered. For the
short range interactions they are infrared finite. The Taylor
expansion of the 1PI graphs in the momenta is well defined
at $p=0$ and the generating functional, the 
effective action is quasi-local. In the
presence of long range interactions there is an IR instability,
the expansion in the momentum is ill defined at $p=0$,
the effective action is non-local. Note that the only way we know for
a massive, non-critical effective actions to be non-local
is the condensation or the spontaneous or dynamical symmetry breaking.
Then the long range interactions indicate that the true vacuum of the
theory is different from the one around which the gradient expansion
of effective action was carried out. In fact, the effective action of
a non-critical theory will always be quasi-local when its field variable
corresponds to the fluctuations around the true vacuum.

The formal indication of long range interactions for the charged particles
in the presence of quenched impurities is that the mean-field 
of the models \cite{elkh}-\cite{beki} is a bilocal field which is proportional 
to the sign of the Matsubara frequency. The mean field generation is
not a spontaneous symmetry breaking, a phenomenon driven by
the infrared modes and leading to homogeneous condensate corresponding
to a local field variable. Instead the non-trivial frequency dependence of
the mean-filed suggests a dynamical symmetry breaking generated by the time
derivative term in the action. The effective theory sought should 
\begin{itemize}
\item be {\bf quasi-local} at the cutoff because the 
relevant part of the interactions are well known in the UV region,
leaving the irrelevant, higher derivative, radiative correction part 
open for phenomenological adjustments, and 
\item contain the {\bf long range interactions} in the IR sector accounting
for the condensate with this particular time dependence. 
\end{itemize}

The solution is the use of higher order derivative terms in time 
which are irrelevant on trivial or on homogeneous background but
can change the universality class when the vacuum is inhomogeneous
\cite{ketd}, \cite{afvac}. In order to keep track of the issue of
locality the effective action will be constructed in time
instead of frequency space.

This program consists of the construction of a chain of effective
theories. We follow the general approach leading to the effective theories
of refs. \cite{elkh}-\cite{beki} except that the auxiliary composite fields,
responsible for the particle-hole and the particle-particle channels will
be kept local instead of bilocal. Such a limitation in the treatment of  non-local 
effects will be compensated for by retaining the terms in the effective action
which are higher order in the time derivative.

The first level is for the electrons and the photons,
represented by the field $\psi$ and the time component of the photon field $u$,
respectively,
\be
Z=\int D[\psi^*]D[\psi]D[u]e^{{i\over\hbar}S[\psi^*,\psi,u]},
\label{secef}
\ee
after having eliminated the quenched impurities by means of the replica 
method. In order to accommodate electrons with finite density 
the chemical potential $\mu$ of the electrons should be introduced in $S$. 
The effective coupling constants in $S[\psi^*,\psi,u]$,
i.e. any interaction vertex beyond the minimal coupling
originate from the interactions between the electrons and the
impurities. This effective theory is supposed to be obtained 
by means of a perturbation expansion whose small parameter is an
effective electron-impurity coupling constant, denoted by $g$.
The Coulomb interaction is kept explicitly in the model by
retaining the field $u$. The reason for eliminating the 
impurities but keeping photons is that the condensation
of charged particles usually induces a mean field, i.e.
condensate for the photon, as well, a phenomenon which is difficult to
reproduce once the photons have already been eliminated perturbatively.
Thus we prefer to keep the photon field at hand to reproduce later 
the condensation in the framework of the saddle point expansion
if needed. The photon dynamics is kept on the tree-level in this paper,
the inclusion of the Coulomb interaction between the electrons will be
ignored. According to the general remarks about non-locality
in time the most important pieces of $S[\psi^*,\psi,u]$ for our purpose
are the terms with higher order derivatives in time.

The influence of quenched impurities on the electrons is usually
taken into account by a static potential, averaged after the quantum
expectation values have already been computed \cite{edwa}. The electron
propagation is influenced for an arbitrary long time by each static, random potential. 
Such a long time modification survives the averaging
over the different realizations of the potential. One can understand this by
recalling that the elimination of a particle always generates non-local 
interactions. The range of this interaction is the correlation length which 
is supposed to be generated by the interaction with the particle in question. 
Since the static potential is of infinitely long range in time its
elimination produces long time correlations. On a more formal level,
the static potential, $V(x)$ multiplies the time integral 
$I(x)=\int dt\psi^*(t,x)\psi(t,x)$ and its fluctuations generate the terms
$I^n(x)$, $n=2,3,\cdots$ in the effective action which are highly non-local 
in time. This non-locality leads to the rather singular 
\be\label{bekimf}
\Lambda(t_1,t_2)\approx\sum_{n>0}^N\sin(t_1-t_2)\pi(2n+1)T
\ee
bilocal mean-field of ref. \cite{elkh}-\cite{beki}, where $T$ is the
temperature. The non-triviality of
this field configuration is due to the non-locality confined
at the ultraviolet cutoff scale, $t_1-t_2\approx1/N$. The extended structure
within this short time interval generates higher order derivatives for the
electron field in the effective action, a generic term for two operators
$A(t)$ and $B(t)$ being
\be
\int dt_1dt_2A(t_1)\Lambda(t_1-t_2)B(t_2)=\sum_{\ell=0}^\infty{G_\ell\over\ell!}
\int dtA(t)\partial_t^\ell B(t),
\ee
where the coupling constant
\be
G_\ell=\int dt\Lambda(t)(-t)^\ell\approx\sum_{n>0}^N{1\over(n\pi T)^{n+1}}
\int_0^{2n\pi}ds(-s)^n\sin s.
\ee
The contribution in the last expression of a given $n$ is
$\ord{(n!/n^{n+1})}$ which gives a rapidly converging series. Thus there will
be few, cutoff independent, higher order derivative terms in the effective
action. This heuristic argument is to explain the relation between the
mean-field \eq{bekimf} and the appearence of the higher order derivatives
only. Instead of a more precise derivation of these effective coupling
constants in the effective theory \eq{secef} we proceed with a more phenomenological 
approach and consider two simple cases only.

Anticipating the importance of composite quasiparticles we 
introduce a $Q_{\alpha,\beta}^{j,k}(x,t)$ charged cooperon and a 
$N_{\alpha,\beta}^{j,k}(x,t)$ ($N^\dagger=N$) neutral diffuson field, 
where $\alpha,\beta$ and $j,k$ are the spin and the replica indices, 
respectively, and approximate \eq{secef} by
\be\label{effth}
Z=\int D[\psi^*]D[\psi]D[Q^*]D[Q]D[N]D[u]
e^{{i\over\hbar}(S_C[u]+S_\psi[\psi^*,\psi,Q^*,Q,N,u]
+\tilde S_{Q,N}[Q^*,Q,N,u])}
\ee
where
\bea\label{effthi}
S_C&=&\int dtdx\hf(\nabla u)^2,\nonu
S_\psi&=&\int dtdx\biggl\{
Z_\psi\psi^{j*}_\alpha K_\psi(i\hbar D_{\psi,t})\psi_\alpha^j
-{\hbar^2\over2m}\psi^{j*}_\alpha\Delta\psi^j_\alpha
-\mu\psi^{*j}_\alpha\psi^{*j}_\alpha\nonu
&&+\psi^{*j}_\alpha\psi^{*k}_\beta Q_{\beta,\alpha}^{k,j}
+Q^{k,j*}_{\beta,\alpha}\psi_\alpha^j\psi_\beta^k
+\psi^{j*}_\alpha N_{\alpha,\beta}^{j,k}\psi_\beta^k\Biggr\},\\
\tilde S_{Q,N}&=&\int dtdx\Biggl\{tr\Biggl
[\tilde Z_QQ^\dagger\tilde K_Q(i\hbar D_{Q,t})Q
-{\hbar^2\over2\tilde M_Q}Q^\dagger\Delta Q\nonu
&&+\tilde Z_NN\tilde K_N(i\hbar\partial_t)N-{\hbar^2\over2\tilde M_N}
N\Delta N\Biggr]+\tilde U\biggr\},\nonumber
\eea
and
\be
D_{\psi,t}=\partial_t-i{e\over\hbar c}u,~~~~
D_{Q,t}=\partial_t-2i{e\over\hbar c}u.
\ee
This effective theory is constructed  in the 
spirit of the Landau-Ginzburg double expansion in the amplitude and the
gradient of the field. The original effective theory \eq{secef} 
contained higher order derivative terms both in time and the space. 
We believe that the non-locality in space is not essential in the 
problem considered and we keep the terms $\ord{(\Delta)}$ only. 
We expect the photon dynamics to be left untouched by the 
localisation thus the $u$ dependence has to be retained in the leading order
and the higher order derivatives terms for the photon field are ignored. 

According to the gradient expansion the coefficient functions
of the retained derivative pieces depend on the 
local field variables, except $u$. The functions 
\be\label{zzz}
\tilde Z_\psi(Q^*,Q,N)=1+O(g),~~\tilde Z_Q(Q^*,Q,N)=O(g),
~~\tilde Z_N(Q^*,Q,N)=O(g),
\ee
and
\be
K_\psi(z)=z+O(g),~~\tilde K_Q(z)=z+O(g),~~\tilde K_N(z)=z+O(g),
\ee
control the strength and the structure of the correlations in time.
These functions will be chosen to be polynomials of finite order 
in order to preserve the quasi-locality at the cutoff of this
effective theory. Note that the time reversal invariance of the original,
microscopic dynamics renders the lagrangian real, in particular
\be
K(z)=\sum\limits_{n=1}^M c_nz^n,
\ee
where the coefficients $c_n$ are real.
The kinetic energy contains the effective masses
\be\label{mmm}
m(Q^*,Q,N)=m+O(g),~~\tilde M_Q^{-1}(Q^*,Q,N)=O(g),
~~\tilde M_N^{-1}(Q^*,Q,N)=O(g).
\ee
$\tilde Z_Q\not=0$ or $\tilde Z_N\not=0$ indicates the presence of bound 
states with diffusion constant $\hbar^2\tilde M_Q^{-1}$,
$\hbar^2\tilde M_N^{-1}$ in the appropriate channels
in the absence of the gap when the fluctuations controled by the
kinetic energy are of long range. The potential $\tilde U(Q^*,Q,N)$
comprises the ultra-local interactions between the composite particles.
The rotational invariance restricts the appearance
of the fields $Q^*$, $Q$, and $N$ in these functions to the
combination $tr(Q^*\sigma_2)^k(\sigma_2Q)^\ell N^m$, where
$\sigma_j$ denotes the Pauli matrices.
The functions above can be obtained by following the standard
procedure of the determination of effective theories \cite{skma}.

Since the long range fluctuations correspond to cooperon or 
diffuson fields the electrons should be eliminated,
\be\label{effqn}
Z=\int D[Q^*]D[Q]D[N]D[u]e^{{i\over\hbar}S_{eff}[Q^*,Q,N,u]}
\ee
where
\bea
S_{eff}&=&\hf trlog
\pmatrix{Q^*&Z_\psi K_\psi-{\hbar^2\over2m}\Delta-\mu+N\cr
Z_\psi K_\psi-{\hbar^2\over2m}\Delta-\mu+N&Q}\\
&&+S_C[u]+\tilde S_{Q,N}[Q^*,Q,N,u].
\eea
By following the same approximation for the fermion determinant
as \eq{effthi} was obtained we arrive at the form
\bea\label{cefac}
S_{eff}[Q^*,Q,N,u]&=&S_C[u]+\int dtdx\Biggl\{tr\Biggl[
Z_QQ^\dagger K_Q(i\hbar D_{Q,t})Q
-{\hbar^2\over2M_Q}Q^\dagger\Delta Q\nonu
&&+Z_NNK_N(i\hbar\partial_t)N
-{\hbar^2\over2M_Q}N\Delta N\Biggr]+U\Biggr\}.
\eea

One could have assumed the form \eq{cefac} as our starting point
and have applied the arguments leading to \eq{zzz}-\eq{mmm} to obtain
directly \eq{cefac} instead of following the sequence 
\eq{secef}$\to$\eq{effth}$\to$\eq{cefac}. The only additional insight
one gains from longer the path is that the auxiliary fields $Q^*$, $Q$ 
and $N$ decouple from the electron field $\psi$ in the weak disorder 
limit, $g\to0$, and a highly singular effective action is found,
\be
e^{{i\over\hbar}\tilde S_{Q,N}[Q^*,Q,N]}\to\delta(Q^*)\delta(Q)\delta(N),
\ee
to suppress their fluctuations.
This problem can be avoided by the introduction of an additional
auxiliary field as in ref. \cite{beki}. Since we are interested
in the dynamics around the phase transition when the disorder
is strong we keep our presentation on the simpler level. 

As mentioned above, the effective theory \eq{cefac} differs from the one proposed
in refs. \cite{elkh}-\cite{beki} by including local fields
but there are higher order derivative terms. The net result of using a quasi-local 
theory is that the saddle point expansion which generates the necessary non-local 
effects is under control because the higher loop contributions remain local,
as opposed to the models with bilocal fields.

It is worthwhile to note the formal similarities between the localization
and the quark confinement. Both are related to the same infrared
instability, the mass gap generation mechanism
and can easily be obtained in local expansions.
The haaron-model for the vacuum of the gluonic sector
offers a view of the confinement as an Anderson localization
in 4+1 dimensions \cite{haar}. Both 
mechanisms are effective at low energies, at high
temperature the quarks deconfine, the electrons delocalize. 
The deconfinement mechanism is related to states with non-zero
coherence length in time which is realized by the condensate of the
gluon field. We shall argue below that the transition
of a strongly disordered system to the conducting phase might be
generated by the coherent states which support the phase coherence in time.
The resulting finite conductivity differs substantially from the classical
one, arising from the incoherent scatterings.

\section{Symmetry breaking}
The effective theory \eq{effqn} with computable but yet unknown
coefficient functions is supposed to generate a long range interaction
in time. Our goal is to show that a suitable chosen 
quasi-local effective theory generates, after a dynamical symmetry
breaking is taken into account, highly non-local interactions.

The key feature, the non-locality is reflected in the higher order 
derivatives and has two important effects. One which comes 
from the perturbation expansion is the appearance of
additional poles on the complex frequency plane
of the loop integrals \cite{poles}. When these poles are at
complex energies then they make the lifetime finite.
The imaginary part of the energy
at the pole casts doubt on the consistency, the unitarity, 
of the model. Perturbation expansion can be used to argue that
the imaginary contributions of the complex poles which always
come into complex conjugate pairs cancel in the optical theorem 
at sufficiently low energy \cite{unit} and such theories can be
consistent \cite{kuti}. Problems may arise at the non-perturbative
level due to the lack of convergent path integral for
the negative norm states corresponding to the complex poles
\cite{unitpr} but reflection positivity \cite{ossc} and the
existence of a subspace with positive definite metric and
unitary time evolution can be proven for a large class 
of models \cite{ketd}. 

Another, non-perturbative
effect of the higher order derivative terms is that they
may generate non-homogeneous vacuum and new critical behavior
\cite{ketd}, \cite{afvac}. The inhomogeneous vacuum in time would
easily be spotted experimentally by the non-conservation of
the energy. Thus the experimentally well confirmed energy
conservation excludes the time dependence of the vacuum.
But the situation changes when the $t\to t+i\tau$ analytic continuation
for imaginary time is considered. In fact, the breakdown of the
translation symmetry of the vacuum in the imaginary time
direction does not imply non-conservation of the energy
for real time processes. In particular, it will be checked that 
the vacuum remains time independent in real time in the
coupling constant range considered. The saddle point
which depends on Im($t$) is not an analytical function of
$t$ thus the Wick rotation becomes rather non-trivial, a complication
whenever the saddle points are available for imaginary
time only. We use in the present work the evolution in imaginary time as a
projection onto the vacuum leaving the issue of the detailed
Wick rotation for subsequent work. Our expressions refer
to a well defined finite Euclidean time interval $\tau$
but in order to find the ground state properties we shall always
be interested in the limit $\tau\to\infty$.

It will be shown that the saddle point of the Euclidean effective
theory \eq{effqn} can be inhomogeneous,
\be\label{inhs}
Q(x,\tau)=\chi(x)e^{i\omega\tau},~~~N(x,\tau)=N_0\cos\omega\tau.
\ee
According to the by now standard semiclassical expansion \cite{raja}
the inhomogeneity of the vacuum gives rise to the dynamical
breakdown of the symmetry with respect to the translations of the
imaginary time, generates Goldstone modes and a
pole in the density-density correlation functions
in the leading order of the saddle point expansion. 
The remarkable effect of the inhomogeneous vacuum in
imaginary time is that some localized electron state may become 
delocalized even at the tree level.
Recall that the single particle states for a given time
are easiest to obtain in the Euclidean theory and the localization
is the result of a destructive interference of the potential
barriers, the capture of the electron states in the 
random, static potential valleys. The (imaginary) time dependence
of $Q(x,\tau)$ not only breaks the time inversion symmetry but
allows the electrons to borrow (Euclidean) energy in units 
of $\hbar\omega$ from the vacuum to escape the
potential valleys and thereby tends to delocalize the system.
Taking into account the inhomogeneous vacuum of the 
appropriate effective theory in the framework of the saddle
point expansion one finds a systematic approach to problems like
the formation of the solid state crystal, the appearance of the
rotons and the periodic ground state for $He^4$, and finally the
dynamic origin of the charge density phase.

We do not attempt to derive here the effective action \eq{effqn}
from an underlying microscopic theory. Instead, we shall be satisfied
to show that for a certain choice of the effective action
the inhomogeneous vacuum \eq{inhs} is formed and the density-density
or the current-current correlation functions contain a pole
and become long-ranged. The latter case is interpreted as the sign of 
the conducting transition in the framework of this effective theory.

\section{Diffusons}
The diffuson denotes a collective mode in the perturbation
expansion which produces an infrared pole in the propagator
of the composite operator $\psi^*\psi\approx N$. We assume that
there are propagating electron-hole states, i.e.
$Z_N\not=0$ and there is a linear term in $K_N(z)$. 
After the rescaling $N\to\sqrt{Z_Nc_1}N$ 
the effective action $S_{Q,N}[Q^*=0,Q=0,N,u=0]$ is 
assumed to contain $Z_N=1$, $M_N=M$, and 
\bea\label{ntrun}
K_N(z)&=&z+c_2z^2+c_3z^3+c_4z^4,\\
U_N(N)&=&\hf G_1tr N^2+\hf G_2(tr N)^2+\hf G_3(trN^2)^2,\nonumber
\eea
where $G_3>0$,
\be
L_N=tr\Biggl[N(i\hbar\partial_t
-c_2\hbar^2\partial_t^2-c_3i\hbar^3\partial_t^3
+c_4\hbar^4\partial_t^4)N
-{\hbar^2\over2M}N\Delta N\Biggr]+U_N(N).
\label{nlagrn}
\ee
The truncation of the potentially infinite order polynomial in 
\eq{ntrun} will be justified later.

The Wick rotation to imaginary time yields
\be
L^E_N=tr\Biggl[N(\hbar\partial_\tau+c_2\hbar^2\partial_\tau^2
+c_3\hbar^3\partial_\tau^3+c_4\hbar^4\partial_\tau^4)N
-{\hbar^2\over2M}N\Delta N\Biggr]+U_N(N)
\ee
where the stability of the vacuum requires the highest order
non-vanishing coefficient, $c_4$, be positive. The path integral
\be
Z=\int D[N]e^{-{1\over\hbar}S^E[N]}
\label{pinn}
\ee
is saturated in the loop expansion by the configuration which
minimizes $ReS^E$ and cancels the derivative of $ImS^E$. A necessary
condition for this is the equation of motion,
\be
[\hbar\partial_\tau+c_2\hbar^2\partial_\tau^2
+c_3\hbar^3\partial_\tau^3+c_4\hbar^4\partial_\tau^4]N
-{\hbar^2\over2M}\Delta N+(G_1+G_3trN^2)N+G_2tr N=0.
\ee
Let us look for the solution of the form
\be
N^{j,k}_{\alpha,\beta}=\delta^{j,k}(n_0(\tau)+n(\tau)\cdot\sigma)_{\alpha,\beta},
\label{ansn}
\ee
where $n_\mu$ $\mu=0,\cdots,3$ are real and $\sigma^j$ 
are the Pauli matrices. The equation of motion is
\be
0=[\hbar\partial_\tau+c_2\hbar^2\partial_\tau^2
+c_3\hbar^3\partial_\tau^3+c_4\hbar^4\partial_\tau^4
+G_1+2G_3n^2+2G_2\delta_{\mu,0}]n_\mu,
\ee
where $n^2=n_\nu n_\nu$. The solution wil further be simplifed by 
assuming the form
\be\label{dsadlp}
n_\mu(\tau)=n_\mu\cos\omega\tau,
\ee
$n_\mu$ being a constant four vector.
We obtain two separate equations since the odd and the even order
time derivatives generate sines and cosines, respectively.
The coefficients of $\sin\omega\tau$ cancel if
\be
\omega^2={1\over\hbar^2c_3}.
\label{freqc}
\ee
The equation for the terms with $\cos\omega t$ is
\be
n_jn_j={(c_2-c_4\hbar^2\omega^2)\hbar^2\omega^2-G_1-2G_2\over2G_3},~~~n_0=0,
\ee
where $j=1,2,3$. In the case $G_2=0$ we find
\be
n_\mu n_\mu={(c_2-c_4\hbar^2\omega^2)\hbar^2\omega^2-G_1\over2G_3}.
\ee

In real time the frequency spectrum is continuous and
\eq{freqc} is replaced by
\be\label{realtsb}
\omega^2=-{1\over\hbar^2c_3},
\ee
indicating the homogeneity of the real time vacuum and the energy
conservation whenever the Euclidean vacuum is time dependent, $c_3>0$.

Consider first the case of 
\be
c_3>0,~~~(c_2-\hbar^2c_4\omega^2)\hbar^2\omega^2>G_1+2G_2,
\label{condinhc}
\ee
and $G_2\not=0$. The spatial rotations are represented as
$\psi_\alpha\to\Omega_{\alpha,\beta}\psi_\beta$, where 
$\Omega\in SU(2)$. The corresponding transformations of the
diffuson field, $N\to\Omega^\dagger N\Omega$,
\be\label{rot}
n_0+n\cdot\sigma\longrightarrow \Omega^\dagger(n_0+n\cdot\sigma)\Omega,
\ee
leave the lagrangian \eq{nlagrn} invariant. But
the rotational symmetry is broken by the $n\not=0$ spin term and \eq{rot}
gives three broken symmetries of the ground vacuum. The fourth
broken continuous symmetry is the translation in the time,
$\tau\to\tau+\tau_0$. Thus there are four soft modes in this phase which support
the long range density fluctuations. Note that
the time translation invariance is broken in the imaginary time
direction only and is left intact for the real time according to \eq{realtsb}. 
For $G_2=0$ the broken symmetry is larger and includes $O(4)$
instead of the rotation group $O(3)$. 

In the regions where \eq{condinhc} is not valid one is left
with the usual Goldstone mode only which corresponds to the
charge conservation and the electron
diffusion reflects the features known from the
perturbation expansion around the trivial vacuum.

$n^2$ and $\omega$ are the two non-trivial parameters of the vacuum 
and the constants $c_j$ and $G_k$ of the action appear through these 
combinations in the tree-level structure. This explains our truncation 
\eq{ntrun}. In fact, the
omitted higher order terms can only modify these two parameters without
introducing any further tree-level effects. It remains to be seen if the 
radiative corrections generate further relevant coupling constants in 
the effective theory.

\section{Cooperons}
To support the conducting phase one needs 
long range current-current correlation functions. 
We shall identify in this section
the coupling constant region of the effective theory for 
the auxiliary field $Q$ where the inhomogeneous vacuum 
produces a desired long range dynamics. 
We start with the effective model
$S_Q[Q^*,Q,u]=S_{Q,N}[Q^*,Q,N=0,u]$ with the truncations
$Z_Q=1$, $M_Q=M$, $U_Q=0$, and
\be\label{qtrun}
K_Q(z)=r^2-(z^2-r)^2+r^{-2}z(z^2-r)^2.
\ee
Since the vacuum will contain a condensate of charges the
static, external electrostatic potential $u_{cr}(x)$ 
of the crystalline structure will be retained, as well. 

The distinguishing feature of the conductance in this model is
that it appears on the tree level without relying on soft fluctuations. 
This happens because the cooperon field is charged and
the corresponding two-electron states propagate on
the periodic background field $u_{cr}(x)$ of the solid state
crystal. The higher order derivatives in the time 
modify the frequency of the saddle point. When the frequency 
is in a band with extended states of $u_{cr}$ then the saddle 
point is delocalized.  We have an inhomogeneous condensate 
of charged particles and a new conduction mechanism in this case. 

The fluctuating component $v(x,t)$ of the temporal photon field, 
\be
u(x,t)=u_{cr}(x)+v(x,t)
\ee
is responsible for the 
projection into the gauge invariant subspace, identified by Gauss' law.
The invariance under the time dependent gauge transformations
\be
Q(x,t)\longrightarrow e^{iq\alpha(t)}Q(x,t),~~~~
u(x,t)\longrightarrow u(x,t)-\partial_t\alpha,
\label{gtra}
\ee
where $q=2e/\hbar c$, requires that the time derivative always 
appears in the combination $D_t=\partial_t+iqu$. 
We expect higher order derivative terms in the
action which introduce higher order, non-linear terms in the photon
field $u$. There is no reason to find non-linearity in the 
external field of the solid state crystal $u_{cr}$, thus the
second equation in \eq{gtra} will be applied for the fluctuations
of the photon field only and the external field is kept gauge invariant
according to the background field method \cite{backg},
\be
u_{cr}(x)\longrightarrow u_{cr}(x),~~~
v(x,t)\longrightarrow v(x,t)-\partial_t\alpha.
\ee
The effective lagrangian in this background gauge is chosen to be
\bea\label{effq}
L_Q&=&tr\Biggl[Q^\dagger(i\hbar D_t-c_2\hbar^2D_t^2
-c_3\hbar^3iD_t^3+c_4\hbar^4D_t^4+ic_5\hbar^5D_t^5)Q\nonu
&&-{\hbar^2\over2M}Q^\dagger\Delta Q
-{2e\over c}u_{cr}Q^\dagger Q\Biggr]+\hf(\nabla u)^2.
\eea

We look now for the possibility of having a non-vanishing conductivity 
produced by the two-electron bound states due to a saddle point in the
 Euclidean theory, 
\be
Q_{cl}(x,\tau)=e^{i\omega\tau}\chi(x),~~~~v_{cl}(x,\tau)=v.
\label{anza}
\ee
It is worthwhile noting that this saddle point breaks the invariance
with respect to the gauge transformation with
\be
\alpha(\tau)={\omega\tau\over q}
\ee
in \eq{gtra} which mixes the positive and negative Matsubara modes.
The imaginary time equations of motion for $\chi$ and $v$ are
the equations for the saddle point,
\be
0=\Biggl[\hbar(\partial_\tau+iqv)+c_2\hbar^2\left(
\partial_\tau^2+iqv\partial_\tau+iq\partial_\tau v-q^2v^2\right)+
\cdots+{\hbar^2\over2m}\Delta-u_{cr}\Biggr]Q,
\ee
and
\be
0=\partial_\tau^2v-\partial^2v+
\hbar trQ^*Q+c_2\hbar^2tr\left(Q^*\partial_\tau Q-(\partial_\tau Q^*)Q
+2qivQ^*Q\right)+\cdots,
\ee
respectively. They are simplified by the  ansatz \eq{anza} to
\bea\label{qcond}
0&=&\Biggl[K_Q(i\xi)+{\hbar^2\over2m}\Delta-u_{cr}\Biggr]\chi\nonu
&=&\Biggl[r^2-(\xi^2+r)^2+ir^{-2}\xi(\xi^2+r)^2
+{\hbar^2\over2m}\Delta-u_{cr}\Biggr]\chi,
\eea
and
\bea\label{vcond}
0&=&{d\over d\xi}K_Q(i\xi)\nonu
&=&{d\over d\xi}\Biggl[r^2-(\xi^2+r)^2+ir^{-2}\xi(\xi^2+r)^2\Biggr],
\eea
where $\xi=\omega+qv$.
Eq. \eq{vcond} is solved first for $v$ for a given $\omega$.
The result is substituted into the first equation which
yields a Schr\"odinger-like equation,
\be
\left[-{\hbar^2\over2m}\Delta+u_{cr}\right]\chi=K_Q(i\xi)\chi.
\label{csch}
\ee
This result motivates our particular choice \eq{qtrun}. In order
for $\chi(x)$ to be extended, $\tilde E=K_Q(i\xi)$ must be real.
This requires in Euclidean space-time that the odd powers of 
$z$ cancel in $K_Q(z)$ at the solution 
of \eq{vcond}. $K_Q(z)$ should contain at least a term
$O(z)$, thus we need a polynomial of order 5. The parametrization 
of $K_Q(z)$ was chosen in such a way that $\xi=\sqrt{-r}$ and the 
eigenvalue of the Schr\"odinger equation is real and positive for $r<0$.

The real time equation of motion for the photon field is
\be
0={d\over d\xi}K_Q(\xi).
\ee
It has no solution with real $\xi$ for $-(5/4)^{1/3}<r<0$ indicating the 
stationarity of the vacuum and the energy conservation in real time
in this parameter regime.

We calculate the conductivity by means of the linear response approach.
First we introduce a weak external electric field,
\be
u_{cr}(x)\longrightarrow u_{cr}(x)-\epsilon z
\ee
and then compute the expectation value of the current operator,
\be\label{curr}
J={\hbar\over2im}tr\left(Q^*\partial Q-\partial Q^*Q\right)
\ee
in the leading order of the saddle point expansion.
In this order the expectation value is given by the
saddle point,
\be\label{cprop}
\la Q^*Q\ra=Q^*_{cl}Q_{cl}+\ord(\hbar).
\ee
The electric field dependence of the saddle point can be
obtained by the help of the Rayleigh-Schr\"odinger perturbation expansion 
for \eq{csch},
\be
\chi_k=\chi^{(0)}_k+\epsilon\sum_{n\not=k}
{\la\chi^{(0)}_k|z|\chi^{(0)}_n\ra\over E^{(0)}_n-E^{(0)}_k}
\chi^{(0)}_n+\ord(\epsilon^2),
\ee
where $\chi^{(0)}_n$ are the solutions of \eq{csch},
by assuming a discrete spectrum, a large but finite quantization volume. 
The real part of the D.C. conductivity of the level $\chi_k$ is then given by
\be
Re\sigma={\hbar\over m}Im\sum_{n\not=k}
{\la\chi^{(0)}_k|z|\chi^{(0)}_n\ra\over E^{(0)}_n-E^{(0)}_k}
\int dx\chi^{(0)*}_k(x)\partial_z\chi^{(0)}_n(x).
\label{scc}
\ee

We are now able to discuss the phase structure
from the point of view of the conductance. Note that we always have
a non-trivial saddle point for $r<0$, $\xi=+\sqrt{-r}$,
$\tilde E=r^2$. On the one hand, since $u_{cr}$ is periodic 
one is always in the continuum of the spectrum of \eq{csch}
for $\tilde E>0$. On the other hand, the saddle point 
is trivial for $r>0$. Thus the long range structure of the
background field changes in the vacuum and there is a 
phase transition at $r=r_0=0$.

The eigenvalue in \eq{csch} 
is positive and falls into the continuum for $r<0$ and we
have $K_Q(i\xi)=E^{(0)}_n$ with $n=k$.
The wave functions $\chi^{(0)}_n(x)$ with 
$E^{(0)}_n\approx E^{(0)}_k$ are extended, as well, and
\eq{scc} is non-vanishing.
Since the parameter $r$ comes from an effective theory it
is $\mu$ dependent. The electron density corresponding to
$\mu_{cr}$ where $r(\mu_{cr})=0$ is the conductor-insulator transition point
in this model. Even if the classical conductivity is canceled
by the cooperon pole contribution in the usual trivial vacuum for 
$\mu<\mu_{cr}$ and the electrons are localized, \eq{scc} represents
a contribution which makes the system conducting when $\mu>\mu_{cr}$.

Note that the conductivity \eq{scc} comes from
a tree level effect and requires no soft modes though they
are present because the vacuum breaks the invariance under
spatial rotations and translation of the imaginary time.
One may call $J$ a supercurrent in \eq{curr} because it corresponds
to the particles which make up the condensate. For the usual 
superfluids the condensate is homogeneous 
and does not support classical current. In our case
the positive energy eigenstate of \eq{csch} are scattering states
and yield a new contribution to the conductivity when the propagators
\eq{cprop} are used in the Kubo formula. The origin of such a
conductivity is the highly populated extended state in Euclidean 
space-time which is generated by the higher order derivative terms 
in the effective action and make the hopping between the adjacent
ions possible.

\section{Summary}
An effective theory was suggested for the description of the
conductor-insulator transition in strongly disordered systems
along the lines of ref. \cite{elkh}-\cite{beki}.
Our starting point is the fact that the main effect of quenched
disorder is an effective interaction for the electrons
which is highly non-local in time. We assumed that this
non-locality can be generated by a quasi-local
effective theory for local fields but having few higher order derivatives
in time. The highly non-local interactions in time are
generated in this model by a ``condensation''
mechanism, in a vacuum which has extended saddle point structure in 
imaginary time.

A spatially homogeneous but imaginary time-dependent neutral
condensate gives rise to Goldstone modes and a pole in the density-density
correlation function in the leading order of the loop
expansion. It remains to be seen if this pole contribution
is canceled by the higher order radiative corrections
as in a homogeneous vacuum \cite{vowo}.

The effective theory supports a space and (imaginary)time dependent
vacuum for certain values of the coupling constants. This vacuum
is a condensate of charged particle-particle modes in an extended 
state and the classical current induced by an external electric field is
non-vanishing. This provides a new conduction mechanism.

The inhomogeneous saddle points are excluded in 1+1 space-time
dimensions and systems in 1 spatial dimensional cannot
acquire a conductive phase by this mechanism \cite{mewa}. But 2 spatial
dimensions allow the dynamical breakdown of continuous
symmetries and a delocalized phase appears.

The effective theory studied here was chosen in such a manner
that its vacuum became inhomogeneous. Further work is clearly needed 
to decide whether such a rather unusual rearrangement can be 
justified in certain materials by a detailed derivation of the 
effective theory from a more fundamental level.

\section{Acknowledgment}
We thank Janos Hajdu for encouragement and useful discussions.
The work was supported in part by the grant OTKA T29927/98
of the Hungarian Academy of Sciences.


\begin{references}
\bibitem{blwi} F. Bloch, \Journal{\ZP}{57}{545}{1929};
A. H. Wilson, \Journal{\PRSLA}{133}{458}{1931}.
\bibitem{peie} R. E. Peierls, {\em Quantum Theory of Solids},
Clarendon press, Oxford, 1955.
\bibitem{andl} P. W. Anderson, \Journal{\PR}{109}{1492}{1958}.
\bibitem{scal} E. Abrahams, P. W. Anderson, D. C. Licciardello, 
T. V. Ramakrishnan, \Journal{\PRL}{42}{673}{1979}.
\bibitem{mott} N. F. Mott, {\em Metal-Insulator transitions}, 
Taylor and Francis, London, 1990.
\bibitem{kolu} W. Kohn, J. M. Luttinger, \Journal{\PR}{108}{590}{1957}.
\bibitem{edwa} S. F. Edwards, \Journal{\PM}{3}{1020}{1958}.
\bibitem{agd} A. A. Abrikosov, L. P. Gorkov, I. E. Dzhyaloshinskii, 
{\em Methods of Quantum Field Theory in Statistical Physics},
Prentice Hall, 1963.
\bibitem{lane} J. S. Langer, T. Neal, \Journal{\PRL}{16}{984}{1966}.
\bibitem{aar} P. W. Anderson, E. Abrahams, T. V. Ramakrishnan,
\Journal{\PRL}{43}{718}{1979}.
\bibitem{glk} L. P. Gorkov, A. I. Larkin, D. E. Khmelnitskii,
\Journal{|JETP}{30}{248}{1979}.
\bibitem{vowo} D. Vollhardt, P. W\"olfle, \Journal{\PRB}{22}{4666}{1980}.
\bibitem{inte} P. A. Lee, T. V. Ramakrishnan, \Journal{\RMP}{57}{287}{1985};
D. Belitz, T. R. Kirkpatrick, \Journal{\RMP}{66}{261}{1994}.
\bibitem{wegn} F. J. Wegner, \Journal{\PRB}{19}{783}{1979}.
\bibitem{scwe} L. Sch\"afer, F. Wegner, \Journal{\ZPB}{38}{113}{1980}.
\bibitem{mcst} A. J. McKane, M. Stone; \Journal{\AP}{131}{36}{1981}.
\bibitem{prsc} A. Pruisken, L. Sch\"afer, \Journal{\NPB}{200}{20}{1982}.
\bibitem{elkh} K. B. Efetov, A. I. Larkin, D. E. Khmelnitski,
\Journal{\JETP}{52}{568}{1980};
\bibitem{fink} A. M. Finkelshtein, \Journal{\JETP}{57}{97}{1983}.
\bibitem{beki}D. Belitz, T. R. Kirkpatrick, \Journal{\PRB}{56}{6513}{1997}.
\bibitem{higgs} P. W. Anderson, \Journal{\PR}{130}{439}{1963};
P. W. Higgs, \Journal{\PL}{12}{132}{1964};
F. Englert, R. Brout, \Journal{\PRL}{13}{321}{1964};
G. S. Guralnik, C. R. Hagen, T. W. Kibble, \Journal{\PRL}{13}{585}{1964};
T. W. Kibble, \Journal{\PR}{155}{1554}{1967}.
\bibitem{erg}  C. Wetterich, \Journal{\PLB}{301}{90}{1993};
T. Morris, \Journal{\IJMPB}{9}{2411}{1994};
J. Alexandre, J. Polonyi, {\em Renormalisation group
for the internal space}, submitted to {\em Ann. Phys.}
\bibitem{ketd} M. Dufour, J. Polonyi, 
{\em Periodic vacuum and particles in two dimensions},
to appear {\em Phys. Rev.} {\bf D}.
\bibitem{afvac} J. Fingberg, J. Polonyi, \Journal{\NPB}{486}{315}{1997};
V. Branchina, H. Mohrbach, J. Polonyi,\Journal{\PRD}{60}{45006}{1999}; 
\Journal{\PRD}{60}{45007}{1999}.
\bibitem{skma} S. K. Ma, {\em Modern Theory of Critical Phenomena},
Benjamin, 1982.
\bibitem{haar} K. Johnson, L. Lellouch, J. Polonyi,
\Journal{\NPB}{367}{675}{1991}.
\bibitem{dimtr} S. Coleman, E. Weinberg, \Journal{\PRD}{7}{1888}{1973}.
\bibitem{jack} R. Jackiw, S. Templeton, \Journal{PRD}{23}{2291}{1981}.
\bibitem{apca} T. Appelquist and J. Carazzone, \Journal{\PRD}{11}{2856}{1975};
T. Appelquist, J. Carazzone, H. Kluberg-Stern, M. Roth,
\Journal{\PRL}{36}{768}{1976}, \Journal{\PRL}{36}{1161}{1976};
T. Appelquist and J. Carazzone, \Journal{\NPB}{120}{77}{1977}.
\bibitem{poles}  A. Pais, G. E. Uhlenbeck, \Journal{\PR}{79}{145}{1950}.
\bibitem{unit} T. D. Lee, G. C. Wick, \Journal{\NPB}{9}{209}{1969};
\Journal{\PRD}{3}{1046}{1979}.
\bibitem{kuti} K. Jansen, J. Kuti, C. Liu, \Journal{\PLB}{309}{119}{1993}.
\bibitem{unitpr} D. G. Boulware, D. J. Gross, \Journal{\NPB}{233}{1}{1984}.
\bibitem{ossc} K. Osterwalder, R. Schrader, \Journal{\CMP}{31}{83}{1973};
\bibitem{raja} R. Rajamaran, {\em Solitons and Instantons}, North Holland,
1982.
\bibitem{backg} L. F. Abbot, \Journal{\NPB}{185}{189}{1981}.
\bibitem{mewa} N. D. Mermin, H. Wagner, \Journal{\PRL}{17}{1133}{1966};
S. Coleman, \Journal{\CMP}{31}{259}{1973}.
\end{references}
\end{document}